\documentclass[12pt]{article}
\usepackage{amsmath,amssymb,bm}
\usepackage{graphicx}
\usepackage{color}
\usepackage{here}
\usepackage{cite}

\DeclareMathOperator{\Tr}{Tr}

\def\Im{\mathop{\rm Im}\nolimits\,}
\def\({\left(}
\def\){\right)}

\def\a{\alpha}
\def\b{\beta}

\def\d{\delta}

\def\f{\phi}
\def\g{\gamma}
\def\h{\theta}

\def\l{\lambda}

\def\s{\sigma}

\def\w{\omega}
\def\x{\xi}

\def\D{\Delta}

\def\P{\Psi}
\def\S{\Sigma}
\def\W{\Omega}

\def\beq{\begin{eqnarray}}
\def\eeq{\end{eqnarray}}

\def\diag{\mathop{\rm diag}\nolimits}

\def\det{\mathop{\rm det}}
\def\CP{\mathbb{CP}}
\def\RP{\mathbb{RP}}
\def\Rt{\mathbb{R}_t}

\newcommand{\bear}{\begin{array}}  
\newcommand {\eear}{\end{array}}

\newcommand{\ds}{\displaystyle}

\numberwithin{equation}{section}

\begin{document}

{\ }
\vspace{-10mm}

\begin{flushright}
{\bf October 2009}\\[1mm]
{\small RIKEN\,-\,TH\,-\,172}\\[1mm]
{\small UT\,-\,09\,-\,21}
\end{flushright}

\vskip 2cm

\begin{center}
\LARGE

\mbox{\bf Scattering of Giant Magnons in $\mathbb{CP}^3$}


\vskip 2cm
\renewcommand{\thefootnote}{$\alph{footnote}$}

\large
\centerline{\sc
Yasuyuki Hatsuda$^{\dagger,\,}$\footnote{{\tt \,hatsuda@riken.jp}}
\quad and \quad\ 
Hiroaki Tanaka$^{\ddagger,\,}$\footnote{{\tt \,tanaka@hep-th.phys.s.u-tokyo.ac.jp}}}

\vskip 1cm

${}^{\dagger}$\emph{Theoretical Physics Laboratory, RIKEN,\\
Wako, Saitama 351-0198, Japan}

\vspace{0.3cm}

${}^{\ddagger}$\emph{Department of Physics, Faculty of Science, University of Tokyo,\\
Bunkyo-ku, Tokyo 113-0033, Japan}

\end{center}

\vskip 14mm

\centerline{\bf Abstract}

\vskip 6mm

We study classical scattering phase of $\mathbb{CP}^2$ dyonic giant magnons in $\mathbb{R}_t \times \mathbb{CP}^3$.
We construct two-soliton solutions explicitly by the dressing method. Using these solutions, we compute the classical time delays for the scattering of giant magnons, and compare them to boundstate $S$-matrix elements derived from the conjectured AdS$_4$/CFT$_3$ $S$-matrix by Ahn and Nepomechie in the strong coupling limit.
Our result is consistent with the conjectured $S$-matrix.
The dyonic solutions play an essential role in revealing the polarization dependence of scattering phase.

\vspace*{1.0cm}

\vfill
\thispagestyle{empty}
\setcounter{page}{0}
\setcounter{footnote}{0}
\renewcommand{\thefootnote}{\arabic{footnote}}
\newpage

\section{Introduction}\label{sec:intro}
The gauge/gravity duality helps us to understand strongly coupled dynamics of gauge theory.
The AdS/CFT correspondence \cite{Maldacena:1997re, Gubser:1998bc, Witten:1998qj} is a suitable arena to study this duality.
In the last year, Aharony, Bergman, Jafferis and Maldacena (ABJM) proposed a 3-dimensional $\mathcal{N}=6$ superconformal Chern-Simons theory with the $U(N) \times U(N)$ gauge group \cite{Aharony:2008ug}. This theory is considered to describe the low energy effective theory of multiple M2-branes.
The ABJM model has another interesting aspect that it gives us a new example of the AdS/CFT correspondence.
The dual string theory is the type IIA superstring theory on $AdS_4 \times \CP^3$ background.
We refer to this duality as AdS$_4$/CFT$_3$ shortly.

In the well-known duality between the four-dimensional ${\cal N}=4$ super Yang-Mills and the type IIB superstrings on $AdS_5 \times S^5$ (AdS$_5$/CFT$_4$ for short), the integrable structure plays a key role in understanding the spectrum of both gauge and string theories.%
\footnote{
On recent developments towards exact spectrum for arbitrary single-trace operators/string states at any coupling, see \cite{Gromov:2009tv, Bombardelli:2009ns, Gromov:2009bc, Arutyunov:2009ur, Frolov:2009in, Hegedus:2009ky, Gromov:2009zb, Roiban:2009aa, Rej:2009dk, Janik:2009dx, Gromov:2009tq}.
}
Amazingly, the AdS$_4$/CFT$_3$ duality also has a similar integrable structure.
On the gauge theory side, the integrability of the ABJM model was found at two loops in \cite{Minahan:2008hf, Gaiotto:2008cg, Bak:2008cp, Zwiebel:2009vb, Minahan:2009te} and at four loops \cite{Bak:2009mq, Minahan:2009aq}, and showed that the spectrum of conformal dimensions is mapped to that of an alternating spin-chain model.
Soon after that, assuming full integrability in the planar limit, the all-loop Bethe ansatz equations and exact $S$-matrix were conjectured \cite{Gromov:2008qe, Ahn:2008aa}.
On the string theory side, it was argued that the Green-Schwarz action on the $AdS_4 \times \mathbb{CP}^3$ background with $\kappa$-symmetry partially fixed can be written as a supercoset sigma-model on $OSp(2,2|6)/[SO(3,1)\times U(3)]$,%
\footnote{
Not all possible motions of strings in $AdS_4 \times \CP^3$ are described by the $OSp(2,2|6)/[SO(3,1)\times U(3)]$ coset model as noted in \cite{Arutyunov:2008if}. The complete Green-Schwarz action describing all possible string motions were constructed in \cite{Gomis:2008jt}. The target superspace of the complete Green-Schwarz superstring action is not a supercoset space.
}
whose  classical integrability was also discussed \cite{Arutyunov:2008if, Stefanski:2008ik} (see also \cite{Gomis:2008jt, Chen:2008qq, Bykov:2009jy}).
The one-loop energy shifts of spinning strings were calculated in \cite{McLoughlin:2008ms, Alday:2008ut, Krishnan:2008zs, Gromov:2008fy, McLoughlin:2008he}.
The results, however, seem to conflict with the prediction from the proposed Bethe ansatz equations.
So far, any convincing explanation of such a disagreement has not been known, thus we need more knowledge and data on the AdS$_4$/CFT$_3$ duality.

As mentioned above, the spectral problem of the ABJM model is mapped to that of the alternating spin-chain model. Once we fix the vacuum of the model, then excitations over the vacuum, which are called magnons, are classified by the residual $SU(2|2)$ symmetry.
The asymptotic spectrum of this model contains an infinite tower of boundstates in addition to fundamental magnons.
As shown in \cite{Beisert:2005tm, Beisert:2006qh, Chen:2006gp}, the (centrally extended) $SU(2|2)$ symmetry constrains the form of boundstate energy:
\begin{align}
\epsilon_Q (P)=\sqrt{\frac{Q^2}{4}+4h^2(\l) \sin^2 \( \frac P2 \)}\;,
\label{eq:dispersion-Q}
\end{align} 
where $P$ is the total momentum, and integer $Q$ is the number of magnons.
We should emphasize that the function $h(\l)$ of the 't Hooft coupling $\l$ is not determined by the $SU(2|2)$ symmetry.
The analysis in the Penrose/BMN limit \cite{Nishioka:2008gz, Gaiotto:2008cg, Grignani:2008is} reveals the leading behavior of $h(\l)$  at strong/weak coupling,%
\footnote{
Recently, the sub-leading term at weak coupling was determined by the four-loop analysis \cite{Minahan:2009aq}: $h(\l)=\l+(-8+\pi^2/3) \l^3+\cdots$.
}
\begin{align}
h(\l)=\begin{cases}
\l+{\cal O}(\l^3)  &(\l \ll 1)\,, \\ 
\ds \sqrt{\frac{\l}{2}}+{\cal O}(1) &(\l \gg 1)\,.
\end{cases}
\end{align}
The string dual of a fundamental magnon excitation is known as giant magnon originally found in \cite{Hofman:2006xt}.
This correspondence is generalized to the one between magnon boundstates and dyonic giant magnons in $\mathbb{R}_t \times S^3$ \cite{Dorey:2006dq, Chen:2006gea}.
In AdS$_4$/CFT$_3$, two kinds of giant magnons are known so far.
One lives in $\CP^1$ subspace of $\CP^3$, and the other in $\RP^2$ \cite{Gaiotto:2008cg, Grignani:2008is, Abbott:2008qd}.
The dyonic generalization of the latter was easily found since $S^3$ dyonic giant magnons can be embedded into the $\RP^3$ subspace. 
The generalization of the former, however, was a difficult problem.
This puzzle was recently solved in a special case \cite{Abbott:2009um} and in a general case \cite{Hollowood:2009sc}.%
\footnote{
Note that the algebraic curve data of this dyonic giant magnon based on \cite{Gromov:2008bz} was found earlier in \cite{Shenderovich:2008bs}.
}
They found dyonic giant magnons in $\CP^2$, which have the same form of dispersion relation \eqref{eq:dispersion-Q}.
In \cite{Hollowood:2009sc}, it was also shown that the dyonic giant magnon in $\RP^3$ is the composite object of two dyonic giant magnons in $\CP^2$.

In the present paper, we study the scattering of two $\CP^2$ dyonic giant magnons.
The scattering matrix is a fundamental tool to explore the AdS/CFT spectrum in the large $R$-charge/spin sector as emphasized in \cite{Staudacher:2004tk}.
The $S$-matrix is also important in studying the finite-size effects via L\"uscher-type corrections in both string and gauge theories \cite{Janik:2007wt, Hatsuda:2008gd, Gromov:2008ie, Heller:2008at, Bajnok:2008bm, Hatsuda:2008na, Bajnok:2008qj, Beccaria:2009eq, Bajnok:2009vm}.
The exact $S$-matrix of the AdS$_4$/CFT$_3$ duality have already been conjectured by Ahn and Nepomechie \cite{Ahn:2008aa}, and passed many non-trivial checks so far \cite{Shenderovich:2008bs, Bombardelli:2008qd, Lukowski:2008eq, Ahn:2008wd, Ahn:2009zg, Zarembo:2009au, Kalousios:2009ey, Sundin:2009zu}.
Since the $\CP^2$ dyonic giant magnons are new solutions, which do not exist in the $AdS_5 \times S^5$ case,
it is an important task to investigate their scattering. It will shed light on understanding of the AdS$_4$/CFT$_3$ duality. 
  
Our goals here are to compute the classical phase shift for the scattering of two $\CP^2$ dyonic magnons, and to compare it to the conjectured $S$-matrix of two magnon boundstates at strong coupling.
To compute the classical phase shift, we first construct general two-soliton solutions of the $\CP^3$ sigma-model. We use the dressing method \cite{Zakharov:1973pp, Zakharov:1980ty, Harnad:1983we}, which is a useful technique to obtain multi-soliton solutions, for the $SU(4)/U(3)$ coset model \cite{Hollowood:2009tw, Kalousios:2009mp, Suzuki:2009sc, Hollowood:2009sc} (see also \cite{Spradlin:2006wk, Kalousios:2006xy} for the AdS$_5$/CFT$_4$ case).
We next read off the classical time delay from these two-soliton solutions.
The phase shift is finally obtained by integrating the time delay with respect to the energy of one of two solitons.
We confirm that our result is consistent with the conjectured AdS$_4$/CFT$_3$ $S$-matrix.
As we will see in Section \ref{sec:2}, there are two kinds of $\CP^2$ dyonic magnons, and
our result shows that these two kinds of giant magnons just correspond to two kinds of excitations in the alternating spin-chain picture: even-site/odd-site excitations.
We also show that our constructed two-soliton solutions reproduce the dyonic giant magnons in $\RP^3$ and ``breather-like" solutions considered in \cite{Hollowood:2009tw, Kalousios:2009mp, Suzuki:2009sc} if choosing the parameters of two solitons appropriately.
This means that these giant magnons can be regarded as composites of two $\CP^2$ dyonic magnons.

The organization of this paper is as follows.
In Section \ref{sec:2}, we give a brief review on the dressing method for the $SU(4)/U(3)$ coset model.
We classify the known one-soliton solutions obtained by this method.
In Section \ref{sec:3}, we construct two-soliton solutions and compute the classical time delays for the scattering of $\CP^2$ dyonic giant magnons.
In Section \ref{sec:comparison}, we compare our result  to the conjectured $S$-matrix.
Section \ref{sec:summary} is devoted to the summary of the paper.
In Appendix \ref{sec:otherGM}, we reproduce other giant magnons than $\CP^2$ ones from our constructed two-soliton solutions.
In Appendix \ref{app:S-bound}, we derive the boundstate $S$-matrix elements by using the fusion procedure.

\section{Dressing Method and Giant Magnons}\label{sec:2}
In this section, we review some fundamental results on the dressing method.
In \cite{Kalousios:2009mp, Suzuki:2009sc, Hollowood:2009tw, Hollowood:2009sc}, one-soliton solutions of the $\CP^3$ sigma-model were considered via dressing method.
In particular, such one-soliton solutions are classified in \cite{Hollowood:2009sc}.
We summarize the classification of one-soliton solutions following \cite{Hollowood:2009sc}.

\subsection{Dressing Method for  $SU(4)/U(3)$ Coset Model}
Our starting point is the two-dimensional sigma-model on $\Rt \times \CP^3$.
Since $\CP^3 \simeq SU(4)/U(3)$,
the sigma model on $\Rt\times\CP^3$ can be formulated in terms of a $SU(4)$-valued field ${\cal F}$.
The embedding coordinate $\bm Z=(Z_1,Z_2,Z_3,Z_4)^t\in \mathbb{C}^4$ of $\CP^3$ is mapped to this $SU(4)$-valued field ${\cal F}$ as
\beq
{\cal F} = \h\left(1-2\frac{\bm{ZZ}^\dagger}{|\bm Z|^2}\right), \label{eq:embed}
\eeq
where
\beq
\h = \diag(-1,1,1,1).
\eeq
Then the bosonic action of the sigma-model is given by
\beq
S =
\int d^2\s \Tr(\partial_a{\cal FF}^{-1}\partial^a{\cal FF}^{-1}),
\label{eq:action}
\eeq
where we fix the worldsheet metric as $h^{ab}=\diag(-1,+1)$.
For later convenience, we rescale the worldsheet coordinates: $(t,x)\equiv(\kappa \tau, \kappa \sigma)$,
and consider the decompactification limit, {\it i.e.}, $\kappa \to \infty$.
Thus $t$ and $x$ run from $-\infty$ to $+\infty$.
We also take the gauge in which the global time of target space is identified as worldsheet time $t$.
The conserved charge corresponding to time translation is given by
\beq
E=\sqrt{\frac{\l}{2}} \int_{-\infty}^\infty dx\,.
\eeq
The equation of motion of \eqref{eq:action} is given by
\beq
\partial_a(\partial^a{\cal FF}^{-1}) = 0 .
\label{eq:EOM}
\eeq
This shows that the current ${\cal J}^a \equiv \partial^a{\cal FF}^{-1}$,
which corresponds to the global symmetry transformation
\beq
{\cal F}\rightarrow U{\cal F}\h U^{-1}\h , \ \ \ U \in SU(4), \label{eq:symtra}
\eeq
is conserved.
The Noether charge is given by
\beq
{\cal Q}_L = \int_{-\infty}^\infty dx \,\partial_t{\cal FF}^{-1}. \label{eq:Noether}
\eeq
In particular, we write the diagonal components of ${\cal Q}_L$ as
\beq
J_k \equiv -i \sqrt{\frac{\l}{2}}({\cal Q}_L )_{kk} \ \ (k=1,2,3,4).\label{eq:angmom}
\eeq
It is easy to check that $J_k$ can be rewritten in terms of the normalized embedding coordinate $\bm z \equiv \bm Z/|\bm Z|$ as
\beq
J_k = 2\sqrt{2\l} \int_{-\infty}^\infty dx \Im (\bar{z}_k D_t z_k)\,,
\eeq
where covariant derivative is defined by $D_a \equiv \partial_a -A_a$ with $A_a = \bar{\bm z} \cdot \partial_a \bm z$.
These conserved charges correspond to angular momenta rotating in $\CP^3$.
Note that they satisfy the equation $\sum_k J_k=0$, thus three of them are independent. 

Now let us proceed to review the dressing method for the $SU(4)/U(3)$ coset model \cite{Kalousios:2009mp, Suzuki:2009sc, Hollowood:2009tw}.
The dressing method is a technique to make a non-trivial solution of equation of motion from a known solution.
The equation of motion \eqref{eq:EOM} is mapped to an equivalent integrability condition
\beq
\partial_{\pm}\P(\l) = \frac{\partial_{\pm}{\cal FF}^{-1}}{1\pm\l}\P(\l), \label{eq:linear}
\eeq
where $x_{\pm}=(t\pm x)/2$ are light-cone coordinates and $\l$ is a complex spectral parameter.
$\cal F$ is the solution of the equation of motion \eqref{eq:EOM}, which is related to the target space coordinate through \eqref{eq:embed}.
$\P(\l)$ is constrained by the SU(4) condition
\beq
\det\P(\l) = 1 , \ \ \ (\P(\bar\l))^\dagger \P(\l) = 1 ,
\eeq
and by the inversion symmetry
\beq
\P(\l) = {\cal F}\h\P(1/\l)\h , \label{eq:inversion}
\eeq
followed from the coset constraint.
When $\l$ equals to zero, we easily find the solution of (\ref{eq:linear}),
\beq
\P(0)={\cal F}.
\eeq
Once solutions ${\cal F}$ and $\P(\l)$ are given, we can construct a new solution of \eqref{eq:linear} by the dressing transformation: $\P'(\l)=\chi(\l)\P(\l)$.
Then the new solution of \eqref{eq:EOM} is obtained by ${\cal F}'=\P'(0)$.
The transformation matrix $\chi(\l)$ is called dressing matrix.

We introduce complex four-component vector $\bm F$ by
\beq
\bm F \equiv \P(\bar\x)\bm \varpi ,
\eeq
where $\bm\varpi$ is a constant complex vector and $\x$ is a constant complex number.
For the $SU(4)/U(3)$ coset model, the dressing matrix is given by
\beq
\chi(\l) = 1 + \frac{Q_1}{\l-\x} + \frac{Q_2}{\l-1/\x}, \label{eq:1qq}
\eeq
where
\beq\begin{split}
Q_1 &= \frac1\D\left[-\frac{\x\bar\x\b}{\x-\bar\x}\bm F\bm F^\dagger
+ \frac{\x\g}{\x\bar\x-1}{\cal F}\h\bm F\bm F^\dagger\right], \\
Q_2 &= \frac1\D\left[\frac\b{\x-\bar\x}{\cal F}\h\bm F\bm F^\dagger\h{\cal F}^\dagger - \frac{\bar\x\g}{\x\bar\x-1}\bm F\bm F^\dagger\h{\cal F}^\dagger\right], \label{eq:q1q2}
\end{split}\eeq
and real numbers $\D,\b,\g$ are defined by
\beq
\b = \bm F^\dagger \bm F, \ \ \ \g = \bm F^\dagger{\cal F}\h\bm F , \ \ \ 
\D = -\frac{\x\bar\x\b^2}{(\x-\bar\x)^2} + \frac{\x\bar\x\g^2}{(\x\bar\x-1)^2}.
\eeq
Multiplying the solution $\Psi(\lambda)$ by the matrix \eqref{eq:1qq}, we obtain the new solution, which is characterized by the parameters $\bm\varpi$, $\x$.
In terms of the projective coordinates, the new solution is expressed by
\beq
\bm Z' = (\tilde\a + \h\bm F\bm F^\dagger \h)\bm Z , \label{eq:newsol}
\eeq
where
\beq
\tilde\a = -\frac{\x\b}{\x-\bar\x}-\frac\g{\x\bar\x-1} .
\eeq
The difference of Noether charge (\ref{eq:Noether}) between old and new solutions can be represented in terms of $Q_1,Q_2$:
\beq
\Delta {\cal Q}_L\equiv{\cal Q}_L^{\rm new}-{\cal Q}_L^{\rm old} = (Q_1+Q_2)\big|_{x=+\infty} - (Q_1+Q_2)\big|_{x=-\infty} . \label{eq:dQbyQ1Q2}
\eeq

We mention the affects of the rotation $\tilde {\bm Z} = U^\dagger \bm Z \ (U\in SU(4))$.
The rotated vectors and matrices are given by,
\beq\begin{split}
\tilde \P(\l) &= \h U^\dagger\h\P(\l)U\,,\quad \tilde{\cal F} = \h U^\dagger\h{\cal F}U\,,\\
\tilde{\bm\varpi} &= U^\dagger\bm\varpi\,,\quad \tilde{\bm F}(\l) =  \h U^\dagger\h\bm F(\l)\,, \quad\tilde\chi(\l) = \h U^\dagger\h\chi(\l)\h U\h.
\end{split}\eeq
This modifies the inversion symmetry (\ref{eq:inversion}) as,
\beq
\tilde\P(\l) = \tilde{\cal F}\h\tilde\P(1/\l)U^\dagger\h U . \label{eq:inversion_mod}
\eeq
Therefore we can set the rotated inversion symmetry (\ref{eq:inversion_mod}) by choosing arbitrary $U\in SU(4)$.
It is the only modification.
Other conditions are not affected by the rotation and the dressing (\ref{eq:newsol}) is consistent with it. 
\subsection{Dressing the Vacuum -- Giant Magnons}
Let us apply the dressing method to the vacuum solution.
We start by the following (BPS) vacuum solution,
\beq
\bm Z_0 = (e^{it/2},0,0,e^{-it/2})^t ,\label{eq:vac}
\eeq
and
\beq
{\cal F}_0 = \begin{pmatrix}
0 & 0 & 0 & e^{it} \\
0 & 1 & 0 & 0 \\
0 & 0 & 1 & 0 \\
-e^{-it} & 0 & 0 & 0
\end{pmatrix} .
\label{eq:vac2}
\eeq
Since the Noether charge of the vacuum solution is easily calculated as
\beq
{\cal Q}_L = i\int_{-\infty}^{\infty} dx \; \diag (1,0,0,-1),
\eeq
the conserved charges satisfy the BPS condition,
\beq
E-\frac{J_1-J_4}{2}=0\,,\quad J_1=-J_4\,,\quad J_2=J_3=0\,.
\eeq
The equation (\ref{eq:linear}) for the vacuum solution \eqref{eq:vac2} is easily solved, and we find
\beq
\P_0(\l) =
\begin{pmatrix}
0 & 0 & 0 & e^{i\S(\l)} \\
0 & 1 & 0 & 0 \\
0 & 0 & 1 & 0 \\
-e^{-i\S(\l)} & 0 & 0 & 0
\end{pmatrix} ,
\label{eq:sol-vac}
\eeq
where
\beq
\S(\l) = \frac{x_+}{1+\l} + \frac{x_-}{1-\l}.
\eeq
The vacuum solution satisfies the rotated inversion symmetry (\ref{eq:inversion_mod}) with $U$ given by (\ref{eq:U}).%
\footnote{It is possible to make the vacuum solution that satisfies the original inversion symmetry (\ref{eq:inversion}), see \cite{Kalousios:2009mp}.}
We multiply the vacuum solution \eqref{eq:sol-vac} by the dressing matrix \eqref{eq:1qq}.
In \cite{Hollowood:2009sc}, the dressed solutions are classified by polarization vector $\bm\varpi=(\varpi_1,\bm\W^t,\varpi_4)^t$.%
\footnote{In this paper, we use the different basis of projective coordinates from that in \cite{Hollowood:2009sc}.
To relate ours with \cite{Hollowood:2009sc}'s, we need to rotate $\bm Z$ and $\bm \varpi$ by the matrix
\beq
U=
\begin{pmatrix}
1/\sqrt2 & 0 & 0 & 1/\sqrt2 \\
0 & 1 & 0 & 0 \\
0 & 0 & 1 & 0 \\
i/\sqrt2 & 0 & 0 & -i/\sqrt2
\end{pmatrix}, \label{eq:U}
\eeq
and to rescale the worldsheet coordinates $(t,x)$ twice.}
There are three types of solutions classified by
\begin{itemize}
\item $\varpi_1 \neq 0$\,, $\varpi_4 = 0$\,,
\item $\varpi_1 = 0$\,, $\varpi_4 \neq 0$\,,
\item $\varpi_1 \neq 0$\,, $\varpi_4 \neq 0$\,.
\end{itemize}
We don't need to consider the case $\varpi_1 = \varpi_4 = 0$ because these solutions are equivalent to the vacuum solution.
In the rest of this section, we construct the dressed solutions for the above three cases, and calculate the conserved charges.

\subsubsection{$\varpi_1 \neq 0$, $\varpi_4 = 0$ case}\label{subsec:w40}
Let us construct the solution for the case $\varpi_1 \neq 0$ and $\varpi_4 = 0$.
Without loss of generality, we can set $\varpi_1=1$.
Using \eqref{eq:newsol}, we obtain the dressed solution,
\beq
\bm Z' = \begin{pmatrix}
\left(-\frac{\x e^u}{\x-\bar\x}-\frac{\bm\W^\dagger\bm\W\bar\x(\x^2-1)}{(\x-\bar\x)(\x\bar\x-1)}\right)e^{it/2} \\
-\bm\W e^{\frac12(u+iv)} \\
\left(-\frac{\bar\x e^u}{\x-\bar\x}-\frac{\bm\W^\dagger\bm\W\bar\x(\x^2-1)}{(\x-\bar\x)(\x\bar\x-1)}\right)e^{-it/2} \label{eq:1solB}
\end{pmatrix},
\eeq
where we introduced two real variables $u$ and $v$ given by
\beq
u = i(\S(\x) - \S(\bar\x)) ,\;\;\; v = \S(\x) + \S(\bar\x) - t .
\eeq
In terms of $\x=r e^{ip/2}$, these are expressed by
\beq
u=(x\cosh\h-t\sinh\h)\cos\a, \;\;\;v =(t\cosh\h-x\sinh\h)\sin\a, \label{eq:uv}
\eeq
where
\beq
\cot\a = \frac{2r}{1-r^2}\sin\(\frac p2\) ,\;\;\;
\tanh\h = \frac{2r}{1+r^2}\cos\(\frac p2 \).\label{eq:alth}
\eeq
Using (\ref{eq:dQbyQ1Q2}), we obtain the difference of conserved charge
\beq
\D{\cal Q}_L = 2i\left|\sin\(\frac p2\)\right|\begin{pmatrix}
-r^{-1} & 0 & 0 \\
0 & -(r-r^{-1})\frac{\bm\W\bm\W^\dagger}{\bm\W^\dagger\bm\W} & 0 \\
0 & 0 & r
\end{pmatrix}.
\eeq
Thus the angular momenta are given by
\begin{align}
J_1 &=J_1^{(0)} -\frac{\sqrt{2\l}}{r}\left|\sin\(\frac p2\)\right| ,\;\;
J_4 =J_4^{(0)} +\sqrt{2\l}r\left|\sin\(\frac p2\)\right| ,\\
J_\W &= -\sqrt{2\l}\left(r-\frac1r\right)\left|\sin\(\frac p2\)\right| , \label{eq:charge}
\end{align}
where $J_\W$ is the angular momentum in the direction of $(0,\bm\W,0)$, and $J_k^{(0)}$ is the angular momentum of the vacuum solution.
Eliminating $r$, we find the dispersion relation
\beq
E-\frac{J_1-J_4}2 = \sqrt{\frac{J_\W^2}{4} + 2\l\sin^2\(\frac p2\)}\, .\label{eq:dispersion}
\eeq
This dispersion relation has the same form as that of dyonic giant magnons on $\Rt \times S^3$ considered in \cite{Chen:2006gea}.
Since the solution \eqref{eq:1solB} lives in the subspace $\CP^2$ of $\CP^3$, we call it $\CP^2$ dyonic giant magnon.
Note that angular momentum $J_\W$ can be also expressed in terms of $\x$ as
\begin{align}
J_\W=\left| \frac{\sqrt{\l/2}}{i}\( \x+\frac{1}{\x}-\bar{\x}-\frac{1}{\bar{\x}} \) \right|\,,\quad
\frac{\x}{\bar{\x}}=e^{ip}\,.
\label{eq:J_W}
\end{align}

As we will see in Section \ref{sec:comparison}, the spectral parameters $\xi$ and $\bar{\xi}$ should be identified as $X^+$ and $X^-$ in the spin-chain description, respectively.
Ordinary, we require $|X^{\pm}|\geq 1$ because of the positivity of the energy.
Therefore we also require $|\xi|\geq 1$.

\subsubsection{$\varpi_1 = 0$, $\varpi_4 \neq 0$ case}
Next, let us consider $\varpi_1 = 0$ and $\varpi_4 \neq 0$.
As mentioned in \cite{Hollowood:2009sc}, the solution with $\bm\varpi$ and $\x$ is same as the one with $\hat{\bm\varpi}=-\h\bm\varpi$ and $\hat{\x}=1/\x$.
Combining this and the symmetry transformation (\ref{eq:symtra}), we can show that the solution with $\bm\varpi'=-U^{-1}\h U\bm\varpi$ and $\x'=1/\x$\ ($U\in SU(4)$) also gives us the same solution.
If we choose $U$ in ($\ref{eq:U}$), the transformation matrix becomes
\beq
-U^{-1}\h U=\begin{pmatrix}
0 & 0 & 0 & 1 \\
0 & -1& 0 & 0 \\
0 & 0 & -1& 0 \\
1 & 0 & 0 & 0
\end{pmatrix}\,,
\label{eq:trans}
\eeq
thus the transformation exchanges $\varpi_{1}$ and $\varpi_4$.
For example, the solution with $\x$ and $\bm\varpi = (0,\bm\W^t,1)^t$ is mapped to the one with $\x'=1/\x$ and $\bm\varpi' = (1,-\bm\W^t,0)^t$.
We can recycle the result in the previous subsection to construct the solution with $\x$ and $\bm\varpi = (0,\bm\W^t,1)^t$, and find
\beq
\bm Z' = \begin{pmatrix}
\left(\frac{\bar\x e^{-u}}{\x-\bar\x}+\frac{\bm\W^\dagger\bm\W\bar\x(\x^2-1)}{(\x-\bar\x)(\x\bar\x-1)}\right)e^{it/2} \\
\bm\W e^{-\frac12(u+iv)} \\
\left(\frac{\x e^{-u}}{\x-\bar\x}+\frac{\bm\W^\dagger\bm\W\bar\x(\x^2-1)}{(\x-\bar\x)(\x\bar\x-1)}\right)e^{-it/2} \label{eq:1solA}
\end{pmatrix},
\eeq
where $u$ and  $v$ are given by (\ref{eq:uv}).
The condition $|\x|\geq 1$ distinguishes this solution from (\ref{eq:1solB}) in the previous subsection.
The angular momenta are given by
\beq\begin{split}
J_1 &=J_1^{(0)} -\sqrt{2\l}r\left|\sin\frac p2\right| ,\;\;
 J_4 = J_4^{(0)}+\frac{\sqrt{2\l}}r\left|\sin\frac p2\right| ,\\
J_\W &= \sqrt{2\l}\left(r-\frac1r\right)\left|\sin\frac p2\right| ,
\end{split}\eeq
and the dispersion relation takes the same form as (\ref{eq:dispersion}).

These two types of $\CP^2$ dyonic giant magnons have a significant meaning. That is, these two types are expected to correspond to even-site or odd-site excitations\footnote{We choose the vacuum of the spin-chain as \eqref{spin chain vacuum}.} of the alternating spin-chain on the gauge theory side.
By the charge consideration, we can guess the correspondence between the polarizations of the $\CP^2$ dyonic magnons and the (bosonic) excitations of the spin-chain as shown in Table \ref{tab:pol}.
Indeed, we will confirm that the scattering phase-shifts of $\CP^2$ dyonic magnons with various polarizations agree with those of the corresponding spin-chain excitations (up to a gauge dependent term).

\begin{table}[tb]
 \caption{Correspondence between the polarizations of giant magnons and the spin-chain excitations. We assume $r=|\xi|>1$.}
 \begin{center}
  \begin{tabular}{cccc}
    \hline
     Polarization $\bm \varpi$  &  $J_2$  &  $J_3$  &  Corresponding excitation  \\
    \hline
     $(1,1,0,0)^t$  &  $-$  &  $0$  &  $Y_4^\dagger \to Y_2^\dagger$  \\
     $(1,0,1,0)^t$  &  $0$  &  $-$  &  $Y_4^\dagger \to Y_3^\dagger$  \\
     $(0,1,0,1)^t$  &  $+$  &  $0$  &  $Y^1 \to Y^2$  \\
     $(0,0,1,1)^t$  &  $0$  &  $+$  &  $Y^1 \to Y^3$  \\
    \hline
  \end{tabular}
  \label{tab:pol}
 \end{center}
\end{table}

\subsubsection{$\varpi_1 \neq 0$, $\varpi_4 \neq 0$ case} \label{subsec:breather}
In this case, we can set $\bm\varpi=(\varpi,\bm\W^t,1/\varpi)^t$ without loss of generality.
Let us write $\varpi=e^c$ where $c$ is a complex constant parameter, then
\beq
\bm F = \begin{pmatrix}
e^{i\bar\S-c} \\
\bm\W \\
-e^{-i\bar\S+c}
\end{pmatrix},
\eeq
where $\bar\S = \S(\bar\x) = (-iu+v+t)/2$.
We can set $c=0$ because $c$ is absorbed by shifting $u$ and $v$.
We easily obtain the solution,
\beq
\bm Z' = \begin{pmatrix}
\left(-\frac{\x e^u+\bar\x e^{-u}}{\x-\bar\x}+\frac{\x\bar\x e^{iv}+e^{-iv}}{\x\bar\x-1}-\frac{\bm\W^\dagger\bm\W\bar\x(\x^2-1)}{(\x-\bar\x)(\x\bar\x-1)}\right)e^{it/2} \\
-\bm\W (e^{\frac12(u+iv)}+e^{-\frac12(u+iv)}) \\
\left(-\frac{\bar\x e^u+\x e^{-u}}{\x-\bar\x}+\frac{e^{iv}+\x\bar\x e^{-iv}}{\x\bar\x-1}-\frac{\bm\W^\dagger\bm\W\bar\x(\x^2-1)}{(\x-\bar\x)(\x\bar\x-1)}\right)e^{-it/2}
\label{eq:1solAB}
\end{pmatrix}.
\eeq
The conserved charges are given by
\begin{align}
&E-\frac{J_1-J_4}{2}=2\sqrt{2\l}\left(r+\frac1r\right)\left|\sin\frac p2\right|, \\
&J_1=-J_4\,,\quad J_2=J_3=0 .
\end{align}
This solution is the breather-like giant magnon considered in \cite{Hollowood:2009tw, Kalousios:2009mp, Suzuki:2009sc}.
The same solution is also obtained by dressing the vacuum twice.
See Appendix \ref{sec:otherGM} for detail.

There is another type of dyonic giant magnons, which live in $\RP^3$ subspace of $\CP^3$.
In Appendix \ref{sec:otherGM}, we will show that these giant magnons can be constructed by using two-soliton solutions obtained in the next section.
This consequence is natural because the dyonic giant magnon in $\RP^3$ is regarded as a composite of two $\CP^2$ dyonic magnons with equal soliton momenta.

\section{Two-Soliton Solutions and Classical Time Delay}\label{sec:3}
In the previous section, we reviewed the classification of one-soliton solutions obtained by the dressing method.
Here we construct two-soliton solutions and calculate the classical time delays of these solutions.

\subsection{Construction of Two-Soliton Solutions}
The way to construct two-soliton solutions is very simple.
We apply the dressing method to one-soliton solutions. 
Let us start with the one-soliton solution with $\xi_1$ and $\bm \varpi_1$,
\begin{align}
\bm Z'=(\tilde{\alpha}_1+\theta \bm F_1 \bm F_1^\dagger \theta) \bm Z_0\,,
\end{align}
where $\bm F_1=\Psi_0(\bar{\xi}_1) \bm\varpi_1$ and
\begin{align}
\tilde{\alpha}_1=-\frac{\xi_1 \beta_1}{\xi_1-\bar{\xi}_1}-\frac{\gamma_1}{\xi_1 \bar{\xi}_1-1}\,,
\end{align}
with $\beta_1=\bm F_1^\dagger \bm F_1,\, \gamma_1=\bm F_1^\dagger {\cal F}_0 \theta \bm F_1$.
Note that $\Psi_0(\lambda)$ and ${\cal F}_0$ are mapped to $\Psi_1(\lambda)$ and ${\cal F}_1$ respectively by the dressing transformation:
\begin{align}
\Psi_1(\l) = \chi(\l) \Psi_0(\l)\,,\quad {\cal F}_1 = \Psi_1(0)=\chi(0){\cal F}_0\,.
\end{align}
The two-soliton solution is obtained by dressing this one-soliton solution.
The solution is given by
\begin{align}
\bm Z''=(\tilde{\alpha}'_2+\theta \bm F'_2 \bm F_2'^\dagger \theta)\bm Z'\,,
\end{align}
where $\bm F_2'=\Psi_1(\bar{\xi}_2) \bm \varpi_2$ and
\begin{align}
\tilde{\alpha}_2'=-\frac{\xi_2 \beta_2'}{\xi_2-\bar{\xi}_2}-\frac{\gamma_2'}{\xi_2 \bar{\xi}_2-1}\,,
\end{align}
with $\beta_2'=\bm F_2'^\dagger \bm F_2',\, \gamma_2'=\bm F_2'^\dagger {\cal F}_1 \theta \bm F_2'$.
After lengthy computation, we finally arrive at the useful expression,
\beq
\bm Z''=(\tilde\a_{11}\tilde\a_{22}-\tilde\a_{12}\tilde\a_{21}+\tilde\a_{22}\h\bm F_1\bm F_1^\dagger \h + \tilde\a_{11}\h\bm F_2\bm F_2^\dagger \h 
- \tilde\a_{12}\h\bm F_1\bm F_2^\dagger \h - \tilde\a_{21}\h\bm F_2\bm F_1^\dagger \h)\bm Z_0 , \label{eq:2sol}
\eeq
where $\bm F_i = \P_0(\bar\x_i)\bm \varpi_i$ $(i=1,2) $ and
\beq
\tilde\a_{ij} = -\frac{\x_i\b_{ij}}{\x_i-\bar\x_j} - \frac{\g_{ij}}{\x_i\bar\x_j-1} \,,
\eeq
with $\b_{ij} = \bm F_i^\dagger \bm F_j , \ \ \ \g_{ij} = \bm F_i^\dagger{\cal F}_0\h\bm F_j$.
This two-soliton solution is characterized by the parameters $\x_i$ and $\bm\varpi_i$ ($i=1,2$).
Here we focus on the case that both solitons are $\CP^2$ dyonic giant magnons.
Since either the first or fourth component of polarization vector $\bm \varpi$ vanishes for $\CP^2$ dyonic giant magnon,
there are essentially two possibilities: $\bm\varpi_1 = (1,\bm\W_1^t,0)^t$, $\bm\varpi_2 = (1,\bm\W_2^t,0)^t$ or $\bm\varpi_1 = (1,\bm\W_1^t,0)^t$, $\bm\varpi_2 = (0,\bm\W_2^t,1)^t$.

First, let us consider $\bm\varpi_1 = (1,\bm\W_1^t,0)^t$ and $\bm\varpi_2 = (1,\bm\W_2^t,0)^t$.
Using (\ref{eq:2sol}), we obtain the explicit profile,
\begin{align}
Z''_1 = &\left[\frac{\xi_1\xi_2}{(\xi_1-\bar\xi_1)(\xi_2-\bar\xi_2)}\left|\frac{\xi_1-\xi_2}{\xi_1-\bar\xi_2}\right|^2e^{u_1+u_2}\right.\label{eq:2solz1} \\
&\left.+K_{22}\frac{\xi_1e^{u_{1}}}{\xi_1-\bar\xi_1}+K_{11}\frac{\xi_2e^{u_{2}}}{\xi_2-\bar\xi_2}
-K_{12}\frac{\xi_2e^{u_{21}}}{\xi_2-\bar\xi_1}-K_{21}\frac{\xi_1e^{u_{12}}}{\xi_1-\bar\xi_2}+(K_{11}K_{22}-K_{12}K_{21})\right]e^{it/2}, \notag \\
Z''_4 = &\left[\frac{\bar\xi_1\bar\xi_2}{(\xi_1-\bar\xi_1)(\xi_2-\bar\xi_2)}\left|\frac{\xi_1-\xi_2}{\xi_1-\bar\xi_2}\right|^2e^{u_1+u_2}\right. \label{eq:2solz4}\\
&\left.+K_{22}\frac{\bar\xi_1e^{u_{1}}}{\xi_1-\bar\xi_1}+K_{11}\frac{\bar\xi_2e^{u_{2}}}{\xi_2-\bar\xi_2}
-K_{12}\frac{\bar\xi_1e^{u_{21}}}{\xi_2-\bar\xi_1}-K_{21}\frac{\bar\xi_2e^{u_{12}}}{\xi_1-\bar\xi_2}+(K_{11}K_{22}-K_{12}K_{21})\right]e^{-it/2}, \notag  \\
\begin{pmatrix}
Z''_2 \\ Z''_3
\end{pmatrix}
&= \left[\frac{\bar\xi_2}{\xi_2-\bar\xi_2}\frac{\xi_1-\xi_2}{\xi_1-\bar\xi_2}e^{u_2+(u_1+iv_1)/2}+K_{22}e^{(u_1+iv_1)/2}-K_{12}e^{(u_2+iv_2)/2}\right]\bm\Omega_1 + (1\leftrightarrow 2), \label{eq:2solz23}
\end{align}
where $u_{ij} \equiv i(\S(\xi_i)-\S(\bar{\xi}_j)),\, v_{ij} \equiv \S(\xi_i)+\S(\bar{\xi}_j)-t$ ($i,j=1,2$), and
\begin{align}
K_{ij} \equiv \frac{(\xi_i^2-1)\bar{\xi}_j \bm \Omega_i^\dagger \bm \Omega_j}{(\xi_i-\bar{\xi}_j)(\xi_i \bar{\xi}_j-1)}\,.
\end{align}
It is useful to express $u_{ij}$ and $v_{ij}$ in terms of $u_i$ and $v_i$,
\begin{align}
u_{ii}&=u_i\,,\;\;u_{12}=\bar{u}_{21}=\frac{u_1+u_2+i(v_1-v_2)}{2}\,,\\
v_{ii}&=v_i\,,\;\;v_{12}=\bar{v}_{21}=\frac{v_1+v_2-i(u_1-u_2)}{2}\,.
\end{align}

Next we construct the two-soliton solution with $\bm\varpi_1 = (1,\bm\W_1^t,0)^t$ and $\bm\varpi_2 = (0,\bm\W_2^t,1)^t$.
As in the one-soliton case, we obtain the desired solution by replacing $\x_2$ and $\bm\varpi_2$ with $\x'_2=1/\x_2$ and $\bm\varpi'_2=-U^{-1}\theta U\bm\varpi_2$ imposing \eqref{eq:trans}.
The final result is lead to
\begin{align}
Z''_1 = &\left[\frac{\xi_1\bar\xi_2}{(\xi_1-\bar\xi_1)(\xi_2-\bar\xi_2)}\left|\frac{\xi_1\xi_2-1}{\xi_1\bar\xi_2-1}\right|^2e^{u_1-u_2}\right. \label{eq:2solABz1}\\
&\left.+K_{22}\frac{\xi_1e^{u_{1}}}{\xi_1-\bar\xi_1}+K_{11}\frac{\bar\xi_2e^{-u_{2}}}{\xi_2-\bar\xi_2}
+K_{12}\frac{e^{-iv_{21}}}{\xi_2\bar\xi_1-1}+K_{21}\frac{\xi_1\bar\xi_2e^{iv_{12}}}{\xi_1\bar\xi_2-1}+(K_{11}K_{22}-K_{12}K_{21})\right]e^{it/2},  \notag \\
Z''_4 = &\left[\frac{\bar\xi_1\xi_2}{(\xi_1-\bar\xi_1)(\xi_2-\bar\xi_2)}\left|\frac{\xi_1\xi_2-1}{\xi_1\bar\xi_2-1}\right|^2e^{u_1-u_2}\right. \label{eq:2solABz4} \\
&\left.+K_{22}\frac{\bar\xi_1e^{u_{1}}}{\xi_1-\bar\xi_1}+K_{11}\frac{\xi_2e^{-u_{2}}}{\xi_2-\bar\xi_2}
+K_{12}\frac{\xi_2\bar\xi_1e^{-iv_{21}}}{\xi_2\bar\xi_1-1}+K_{21}\frac{e^{iv_{12}}}{\xi_1-\bar\xi_2}+(K_{11}K_{22}-K_{12}K_{21})\right]e^{-it/2}, \notag \\
\begin{pmatrix}
Z''_2 \\ Z''_3
\end{pmatrix}
= &\left[\frac{\bar\xi_2}{\xi_2-\bar\xi_2}\frac{\xi_1\xi_2-1}{\xi_1\bar\xi_2-1}e^{-u_2+(u_1+iv_1)/2}+K_{22}e^{ (u_1+iv_1)/2}-K_{12}e^{-(u_2+iv_2)/2}\right]\bm\Omega_1 \label{eq:2solABz23}\\
 &+\left[\frac{\bar\xi_1}{\xi_1-\bar\xi_1}\frac{\xi_2\xi_1-1}{\xi_2\bar\xi_1-1}e^{ u_1-(u_2+iv_2)/2}+K_{11}e^{-(u_2+iv_2)/2}-K_{21}e^{ (u_1+iv_1)/2}\right]\bm\Omega_2\,. \notag
\end{align}

\subsection{Classical Time Delay for Two-Soliton Scattering}
Since we have obtained the explicit profiles of two-soliton solutions, we can proceed to the stage to evaluate the classical time delay of two-soliton scattering.
Below we consider the time delay when soliton 1 passes soliton 2.
The velocity of soliton $i$ is given by $V_i=\tanh\h_i$ where $\theta_i$ is defined by (\ref{eq:alth}).
We assume $V_1>V_2>0$.
To find the time delay, we substitute $x=V_1(t-\d t)$ into \eqref{eq:2solz1}-\eqref{eq:2solz23} or \eqref{eq:2solABz1}-\eqref{eq:2solABz23}, and compare the behavior in the asymptotic limit $t\rightarrow\pm\infty$ to the one-soliton solution with $x=V_1 t$.

Let us consider the scattering of two $\CP^2$ dyonic giant magnons with $\bm\varpi_1 = (1,\bm\W_1^t,0)^t$ and $\bm\varpi_2 = (1,\bm\W_2^t,0)^t$.
Without loss of generality, we can set
\beq
\bm\varpi_1 = (1,1,0,0)^t, \;\;\;
\bm\varpi_2 = (1,e^{i\alpha}\cos \phi, e^{i\beta} \sin \phi,0)^t.
\eeq
At $t\to -\infty$, the solution \eqref{eq:2solz1}-\eqref{eq:2solz23} reduces to
\begin{align}
Z''_1 &\rightarrow \frac{K_{11}K_{22}-K_{12}K_{21}}{K_{11}}
\left[\left(1-\frac{K_{12}K_{21}}{K_{11}K_{22}}\right)^{-1}\frac{\xi_1e^{u_1}}{\xi_1-\bar\xi_1}+K_{11}\right]e^{it/2}\,, \\
Z''_4 &\rightarrow \frac{K_{11}K_{22}-K_{12}K_{21}}{K_{11}}
\left[\left(1-\frac{K_{12}K_{21}}{K_{11}K_{22}}\right)^{-1}\frac{\bar\xi_1e^{u_1}}{\xi_1-\bar\xi_1}+K_{11}\right]e^{-it/2}\,, \\
\begin{pmatrix}
Z''_2 \\ Z''_3
\end{pmatrix}
&\rightarrow \frac{K_{11}K_{22}-K_{12}K_{21}}{K_{11}}\left[\left(1-\frac{K_{12}K_{21}}{K_{11}K_{22}}\right)^{-1}\begin{pmatrix}
1-\frac{K_{21}}{K_{22}}e^{i\a}\cos\f \\ \frac{K_{21}}{K_{22}}e^{i\b}\sin\f
\end{pmatrix}\right]e^{(u_1+iv_1)/2}\,.
\end{align}
Here
\beq
1-\frac{K_{12}K_{21}}{K_{11}K_{22}}=\left|\frac{(\x_1-\x_2)(\x_1\x_2-1)}{(\x_1-\bar\x_2)(\x_1\bar\x_2-1)}\right|^2\cos^2\f+\sin^2\f\,,
\eeq
then we can read off the time delay easily,
\beq
\d t_- =- i \frac{|\x_1-1|^2|\x_1+1|^2}{\x_1^2-\bar\x_1^2}\log \left[\left|\frac{\x_1-\x_2}{\x_1-\bar\x_2}\cdot \frac{\x_1\x_2-1}{\x_1\bar\x_2-1}\right|^2\cos^2\f+\sin^2\f\right]\,.
\eeq
Similarly, we can compute the behavior at $t\to +\infty$:
\begin{align}
Z''_1 &\rightarrow \frac{\xi_2e^{u_2}}{\xi_2-\bar\xi_2}\left[\left|\frac{\xi_1-\xi_2}{\xi_1-\bar\xi_2}\right|^2\frac{\xi_1e^{u_1}}{\xi_1-\bar\xi_1}+K_{11}\right]e^{it/2}\,, \\
Z''_4 &\rightarrow \frac{\bar\xi_2e^{u_2}}{\xi_2-\bar\xi_2}\left[\left|\frac{\xi_1-\xi_2}{\xi_1-\bar\xi_2}\right|^2\frac{\bar\xi_1e^{u_1}}{\xi_1-\bar\xi_1}+K_{11}\right]e^{-it/2} \,,\\
\begin{pmatrix}
Z''_2 \\ Z''_3
\end{pmatrix}
&\rightarrow \frac{\bar\xi_2e^{u_2}}{\xi_2-\bar\xi_2}\frac{\xi_1-\xi_2}{\xi_1-\bar\xi_2}e^{(u_1+iv_1)/2}\bm \Omega_1\,,
\end{align}
thus the time delay is given by
\beq
\d t_+= i \frac{|\x_1-1|^2|\x_1+1|^2}{\x_1^2-\bar\x_1^2}\log\left|\frac{\x_1-\x_2}{\x_1-\bar\x_2}\right|^2.
\eeq
Combining above results, we get the total time delay,
\beq\begin{split}
\D T_{12}&\equiv \d t_+-\d t_- \\
&= i \frac{|\x_1-1|^2|\x_1+1|^2}{\x_1^2-\bar\x_1^2}\log\left[\left|\frac{\x_1-\x_2}{\x_1-\bar\x_2}\right|^4\left|\frac{\x_1\x_2-1}{\x_1\bar\x_2-1}\right|^2\cos^2\f
+\left|\frac{\x_1-\x_2}{\x_1-\bar\x_2}\right|^2\sin^2\f\right].
\label{eq:time-delay-1}
\end{split}\eeq

For the case $\bm \varpi_1=(1,1,0,0)^t$ and $\bm \varpi_2=(0,e^{i\a}\cos\f,e^{i\b}\sin\f,1)^t$, 
the computation can be done in a similar manner.
We write only the final result,
\beq
\D \tilde{T}_{12} = -i\frac{|\x_1-1|^2|\x_1+1|^2}{\x_1^2-\bar\x_1^2}\log\left[\left|\frac{\x_1-\x_2}{\x_1-\bar\x_2}\right|^2\left|\frac{\x_1\x_2-1}{\x_1\bar\x_2-1}\right|^4\cos^2\f
+\left|\frac{\x_1\x_2-1}{\x_1\bar\x_2-1}\right|^2\sin^2\f\right].
\label{eq:time-delay-2}
\eeq
In particular, the following four types of scattering are important for the comparison with the proposed AdS$_4$/CFT$_3$ $S$-matrix (see Table \ref{tab:pol}):

(I) $\bm \varpi_1=(1,1,0,0)^t$ and $\bm \varpi_2=(1,1,0,0)^t$,
\beq
\Delta T_{12}^{(\rm I)}=2i \frac{|\x_1-1|^2|\x_1+1|^2}{\x_1^2-\bar\x_1^2}\log\left[\left|\frac{\x_1-\x_2}{\x_1-\bar\x_2}\right|^2\left|\frac{\x_1\x_2-1}{\x_1\bar\x_2-1}\right| \right]\,, 
\label{eq:T^I}
\eeq

(II) $\bm \varpi_1=(1,1,0,0)^t$ and  $\bm \varpi_2=(1,0,1,0)^t$, 
\beq
\Delta T_{12}^{(\rm II)}=2i \frac{|\x_1-1|^2|\x_1+1|^2}{\x_1^2-\bar\x_1^2}\log\left[\left|\frac{\x_1-\x_2}{\x_1-\bar\x_2}\right| \right]\,,
\eeq

(III) $\bm \varpi_1=(1,1,0,0)^t$ and $\bm \varpi_2=(0,0,1,1)^t$,
\beq
\Delta T_{12}^{(\rm III)}=-2i \frac{|\x_1-1|^2|\x_1+1|^2}{\x_1^2-\bar\x_1^2}\log\left[\left|\frac{\x_1\x_2-1}{\x_1\bar\x_2-1}\right| \right]\,, 
\eeq

(IV) $\bm \varpi_1=(1,1,0,0)^t$ and $\bm \varpi_2=(0,1,0,1)^t$,
\beq
\Delta T_{12}^{(\rm IV)}=-2i \frac{|\x_1-1|^2|\x_1+1|^2}{\x_1^2-\bar\x_1^2}\log\left[\left|\frac{\x_1-\x_2}{\x_1-\bar\x_2}\right|\left|\frac{\x_1\x_2-1}{\x_1\bar\x_2-1}\right|^2 \right]\,. 
\label{eq:T^IV}
\eeq
To evaluate the classical phase shift, we integrate the time delay with respect to the energy of soliton 1.

In the end of  this section, we notice the scattering of two elementary giant magnons.
If we take the limit $|\xi_i| \to 1$, the $\CP^2$ dyonic giant magnons reduce to the $\CP^1$ giant magnons.%
\footnote{
More precisely, we have to tune $\bm\W_i$ at the same time. See Appendix \ref{sec:otherGM} for detail. 
} 
In this limit, the time delays \eqref{eq:time-delay-1} and \eqref{eq:time-delay-2} do not depend on $\phi$, and become equal.
This means that the polarization dependence of scattering phases cannot be observed as long as we consider the scattering of elementary giant magnons.
We need to consider the dyonic solutions to distinguish the scattering of various polarizations.

\section{Comparison with Proposed $S$-matrix}\label{sec:comparison}
In this section, we compare the classical scattering phase-shift computed in the previous section to the proposed exact $S$-matrix in \cite{Ahn:2008aa}. The ABJM model has four complex scalar fields $Y^i$ ($i=1,\dots,4$).
Here we take the groundstate of the spin-chain with infinite length as the following BPS operator:
\begin{equation}
\Tr [(Y^1 Y_4^\dagger)^L]\,, \qquad (L\to \infty)\,.
\label{spin chain vacuum}
\end{equation} 
The choice of the vacuum preserves the $SU(2|2)$ symmetry of $OSp(2,2|6)$.
This residual symmetry classifies excitations over the vacuum \eqref{spin chain vacuum}.
The excitations transform in short representations of $SU(2|2)$.
Note that there are two types of elementary excitations replacing odd-site $Y^1$ or even-site $Y_4^\dagger$ by fields in the fundamental representation of $SU(2|2)$. 

The two-body $S$-matrix between elementary excitations can be written as
\begin{align}
\mathbb{S}^{OO}(x_1^\pm,x_2^\pm)&=\mathbb{S}^{EE}(x_1^\pm,x_2^\pm)=S_0(x_1^\pm,x_2^\pm) \hat{\mathbb{S}}(x_1^\pm,x_2^\pm)\,, \label{eq:S^OO}\\
\mathbb{S}^{OE}(x_1^\pm,x_2^\pm)&=\mathbb{S}^{EO}(x_1^\pm,x_2^\pm)=\tilde{S}_0(x_1^\pm,x_2^\pm) \hat{\mathbb{S}}(x_1^\pm,x_2^\pm)\,,
\label{eq:S^OE}
\end{align}
where $\mathbb{S}^{OO}(x_1^\pm,x_2^\pm)$ is the $S$-matrix of two odd-site excitations, and so on.
The spectral parameters $x_j^\pm \ (j=1,2)$ satisfy the equations:
\begin{align}
\left( x_j^+ + \frac{1}{x_j^+}\right) - \left( x_j^- +\frac{1}{x_j^-}\right)=\frac{i}{h(\lambda)},\qquad
\frac{x_j^+}{x_j^-}=e^{ip_j}\,,
\end{align}
where $p_j$ is a momentum of each magnon.
The matrix part $\hat{\mathbb{S}}(x_1^\pm,x_2^\pm)$ is determined by the $SU(2|2)$ symmetry, and takes the same form as the case of AdS$_5$/CFT$_4$ \cite{Beisert:2005tm, Beisert:2006qh, Arutyunov:2006yd}.
The scalar parts $S_0(x_1^\pm,x_2^\pm),\tilde{S}_0(x_1^\pm,x_2^\pm)$ are constrained by the unitarity, crossing invariance \cite{Janik:2006dc} and charge conjugation. It is proposed in \cite{Ahn:2008aa} that they should take the form
\begin{align}
S_0(x_1^\pm,x_2^\pm) = \frac{1-1/(x_1^+ x_2^-)}{1-1/(x_1^- x_2^+)} \, \sigma(x_1^\pm,x_2^\pm)\,, \qquad
\tilde{S}_0(x_1^\pm,x_2^\pm) = \frac{x_1^- - x_2^+}{x_1^+ - x_2^-} \, \sigma(x_1^\pm,x_2^\pm)\,,
\label{eq:scalar-part}
\end{align}
where $\sigma (x_1^\pm,x_2^\pm)$ is the BHL/BES dressing phase \cite{Beisert:2006ib, Beisert:2006ez} in which $g_{AdS_5} \equiv \sqrt{\lambda_{AdS_5}}/(4\pi)$ is replaced by $h(\lambda)$.
 
In order to compare the result in the previous section to the proposed $S$-matrix, we have to consider the scattering of two BPS boundstates.
According to Table \ref{tab:pol}, the $\CP^2$ dyonic magnon with $\bm \varpi=(1,1,0,0)^t$ corresponds to the magnon boundstate such that all constituents are even-site excitations $\phi_E^1(=Y_2^\dagger)$, for example.%
\footnote{We denote  the fundamental fields of $SU(2|2)$ as $\{\phi^1,\phi^2|\psi^1,\psi^2\}$.}
Thus, it suffices to consider the following four types of boundstate scattering:

(I) \;\; $\phi_E^1(X_1^\pm)\phi_E^1(X_2^\pm) \to \phi_E^1(X_1^\pm)\phi_E^1(X_2^\pm)$,

(II) \, $\phi_E^1(X_1^\pm)\phi_E^2(X_2^\pm) \to \phi_E^1(X_1^\pm)\phi_E^2(X_2^\pm)$,

(III) $\phi_E^1(X_1^\pm)\phi_O^1(X_2^\pm) \to \phi_E^1(X_1^\pm)\phi_O^1(X_2^\pm)$,

(IV) $\phi_E^1(X_1^\pm)\phi_O^2(X_2^\pm) \to \phi_E^1(X_1^\pm)\phi_O^2(X_2^\pm)$,\\
where $\phi_{E/O}^i(X^\pm)$ denotes the boundstate with spectral parameters $X^\pm$ consisting of all even/odd-site excitations $\phi_{E/O}^i$.
These $S$-matrices can be obtained by using the fusion procedure.%
\footnote{Note that the full $SU(2|2)$-invariant $S$-matrix for boundstates was obtained in \cite{Arutyunov:2009mi} based on the formulation of \cite{Arutyunov:2008zt}.
Their result, however, is complicated, and we could not read off our interested components of the full $S$-matrix. For some special cases, we can easily construct the boundstate $S$-matrix elements by the fusion procedure as done here.} 
In Appendix \ref{app:S-bound}, we derive them in detail.
The results are given by
\begin{align}
S^{(\rm I)}(X_1^\pm,X_2^\pm)&=S_{\rm BDS}(X_1^\pm,X_2^\pm) \sigma(X_1^\pm,X_2^\pm)\,, \label{eq:S^I}\\
S^{(\rm II)}(X_1^\pm,X_2^\pm)&=S_{\rm BDS}(X_1^\pm,X_2^\pm) S_2(X_1^\pm,X_2^\pm) \sigma(X_1^\pm,X_2^\pm)\,,\\
S^{(\rm III)}(X_1^\pm,X_2^\pm)&=\sigma(X_1^\pm,X_2^\pm)\,,\\
S^{(\rm IV)}(X_1^\pm,X_2^\pm)&=S_2(X_1^\pm,X_2^\pm) \sigma(X_1^\pm,X_2^\pm)\,, \label{eq:S^IV}
\end{align}
where $S_{\rm BDS}(X_1^\pm,X_2^\pm)$ and $S_2(X_1^\pm,X_2^\pm)$ take the forms \eqref{eq:S_BDS} and \eqref{eq:S_2}, respectively.
The spectral parameters of boundstates satisfy the relations,
\begin{align}
\left( X_j^+ + \frac{1}{X_j^+}\right) - \left( X_j^- +\frac{1}{X_j^-}\right)=\frac{i Q_j}{h(\lambda)},\qquad
\frac{X_j^+}{X_j^-}=e^{iP_j}\,,\;\;\;(j=1,2)\,,
\label{eq:boundstate-Q}
\end{align}
where $P_1,\,P_2$ are the total momenta of boundstates.
The energies of boundstates are  given by
\begin{align}
E_j=\frac{h(\lambda)}{2i} \left[ \left( X_j^+ - \frac{1}{X_j^+}\right) - \left( X_j^- -\frac{1}{X_j^-}\right)\right]=\sqrt{\frac{Q_j^2}{4}+4h^2(\lambda) \sin^2 \left(\frac{P_j}{2} \right)}\,.
\end{align}
Comparing \eqref{eq:boundstate-Q} to \eqref{eq:J_W}, we should identify the spectral parameters here with those in the dressing method:
\begin{align}
X_1^+ = \xi_1\,,\;\; X_1^-=\bar{\xi}_1\,,\;\;X_2^+ = \xi_2\,,\;\; X_2^-=\bar{\xi}_2\,.
\label{eq:ident}
\end{align}

Now let us consider the strong coupling limit of the $S$-matrices \eqref{eq:S^I}-\eqref{eq:S^IV}.
In the strong coupling limit, the leading contribution of the BDS part is given by
\begin{align} 
S_{\rm BDS}(X_1^\pm,X_2^\pm) \approx \exp [ -i \Theta_{\rm BDS}(X_1^\pm,X_2^\pm)]\,,
\end{align} 
where
\begin{align}
\Theta_{\rm BDS}(X_1^\pm,X_2^\pm)&=-2h(\lambda) [ \hat{k}(X_1^+,X_2^+)+\hat{k}(X_1^-,X_2^-)-\hat{k}(X_1^+,X_2^-)-\hat{k}(X_1^-,X_2^+)]\,, \\
\hat{k}(X,Y)&=\left[ \left( X+\frac{1}{X} \right) - \left( Y+\frac{1}{Y} \right) \right]
\log \left[ (X-Y)\left(1-\frac{1}{XY}\right)\right]\,.
\end{align}
It is easy to check that $S_2(X_1^\pm,X_2^\pm) \approx \exp [ -i \Theta_{2}(X_1^\pm,X_2^\pm)]$ behaves as
\beq
\Theta_{2}(X_1^\pm,X_2^\pm)=-\frac{1}{2} \Theta_{\rm BDS}(X_1^\pm,X_2^\pm)\,.
\eeq
Finally the dressing part takes the form
\begin{align}
\sigma(X_1^\pm,X_2^\pm) \approx \exp[- i \Theta_{d}(X_1^\pm,X_2^\pm) ]\,,
\end{align}
where
\begin{align}
\Theta_{d}(X_1^\pm,X_2^\pm)&=-h(\lambda)[ k_0(X_1^+,X_2^+)+k_0(X_1^-,X_2^-)-k_0(X_1^+,X_2^-)-k_0(X_1^-,X_2^+)]\,, \\
k_0(X,Y) &\equiv -\left[ \left( X+\frac{1}{X} \right) - \left( Y+\frac{1}{Y} \right) \right]
\log \left(1-\frac{1}{XY}\right)\,.
\end{align}
Collecting these results, we obtain the strong coupling limit of \eqref{eq:S^I}-\eqref{eq:S^IV},
\begin{align}
S^{(X)}(X_1^\pm,X_2^\pm) &\approx \exp [-i \Theta_{\text{spin-chain}}^{(X)} (X_1^\pm, X_2^\pm) ]\;\;\;\; (X={\rm I, II, III, IV})\,,  
\end{align}
where
\begin{alignat}{2}
\Theta_{\text{spin-chain}}^{(\rm I)}&= \Theta_{\rm BDS}+\Theta_{d}\,, &\qquad 
\Theta_{\text{spin-chain}}^{(\rm II)}&= \Theta_{\rm BDS}+\Theta_2+\Theta_{d} \,, \\
\Theta_{\text{spin-chain}}^{(\rm III)}&= \Theta_{d}\,,& 
\Theta_{\text{spin-chain}}^{(\rm IV)}&= \Theta_{2}+\Theta_{d}\,.
\end{alignat}
The time delay is given by the derivative of the scattering phase with respect to the energy of boundstate $X_1^\pm$. 
Using the identification \eqref{eq:ident}, we find
\begin{align}
\Delta \tau_{12}^{(X)} \equiv \frac{\partial \Theta^{(X)}_{\text{spin-chain}}}{\partial E_1}=\Delta T^{(X)}_{12}+i\left(\frac{1}{\xi_2}-\frac{1}{\bar{\xi}_2}\right) \frac{\xi_1 \bar{\xi}_1+1}{\xi_1 +\bar{\xi}_1}\;\;\; (X={\rm I, II, III, IV})\,,
\end{align}
where $\Delta T_{12}^{(X)} \;\;(X={\rm I, II, III, IV})$ are the time delays of $\CP^2$ dyonic giant magnons given by \eqref{eq:T^I}-\eqref{eq:T^IV}.
These results lead to the relations of phase-shifts on both sides,
\begin{align}
\Theta^{(X)}_{\text{spin-chain}}&=\Theta^{(X)}_{\text{string}}+(2E_2-Q_2)P_1\;\;\;(X={\rm I, II, III, IV})\,,  \label{eq:phase-1}
\end{align}
As pointed out in \cite{Hofman:2006xt, Chen:2006gq}, the second term above arises due to the difference of gauge choices in string and gauge theories.
On the string theory side, we choose the gauge such that the density of energy $E$ is constant.
On the gauge theory side, we choose the gauge \eqref{eq:sc-frame} such that a unit length of spin-chain is assigned to a field in the single-trace operator.
Because of such a difference, the spin-chain length and the worldsheet length are also different.
In the end, the $S$-matrix in different gauges differs by an extra factor. 
See \cite{Hofman:2006xt, Chen:2006gq} for more detail. 

\section{Summary}\label{sec:summary}
In this paper, we consider the classical scattering of two $\CP^2$ dyonic giant magnons, which are fundamental objects in the AdS$_4$/CFT$_3$ duality.
Using the dressing method for the $SU(4)/U(3)$ coset model, we constructed general two-soliton solutions on $\mathbb{R}_t \times \CP^3$.
These solutions allow us to compute the classical time delay of the soliton scattering.
We compared the time delay to the proposed $S$-matrix in the ABJM model.

Our computation shows that the classical scattering phase of $\CP^2$ dyonic giant magnons agrees with that of magnon boundstates in the ABJM model up to the gauge dependent term.
In the ABJM model, there are two types of excitations, that is, odd-site excitations and even-site excitations.
The correspondence between the polarizations of giant magnons and the spin-chain excitations is summarized in Table \ref{tab:pol}. 
The polarization dependence of scattering phase does not appear as long as we consider the scattering of elementary (giant) magnons, and thus we need to consider dyonic solutions/magnon boundstates to see it. 

There are some directions to be studied in the future.
It is interesting to consider the one-loop correction of the scattering phase as done in the AdS$_5$/CFT$_4$ case \cite{Chen:2007vs}. 
For this purpose, we need to analyze the quantum fluctuations around the classical background (giant magnon).
The classical scattering data for small fluctuations around the giant magnon can be obtained from multi-soliton solutions.
In the AdS$_4$/CFT$_3$ case, however, one should be careful that the sub-leading term of $h(\l)$ might also contribute to the phase-shift at one-loop level.
Thus it is an important task to know the sub-leading behavior of $h(\l)$ at strong coupling.
It is also interesting to investigate the scattering of spiky strings.
In \cite{Ishizeki:2007kh}, it was shown that the classical phase-shift of two single spike solutions is curiously the same as that of giant magnons up to non-logarithmic terms. The gauge theory interpretation of the single spike solution is unclear so far.
For the AdS$_4$/CFT$_3$ case, it is intriguing to compute the phase-shift of single spikes and to confirm whether it is also same as that of giant magnons or not.
Our two-soliton solutions would be useful for this purpose.


\subsubsection*{Acknowledgments}

We would like to thank K. Hashimoto, K. Okamura and R. Suzuki for reading the manuscript and giving us valuable comments.
YH is also grateful to T.-S. Tai and W.-Y. Wen for useful discussions.
The work of YH is supported in part by JSPS Research Fellowships for Young Scientists.
HT is supported by Global COE Program ``the Physical Sciences Frontier", MEXT, Japan.

\appendix
\section{Other Giant Magnons}\label{sec:otherGM}
Here we show that $\CP^2$ dyonic giant magnons produce other types of giant magnons.
We also consider the limit in which dyonic solutions reduce to elementary giant magnons.

\subsection{Dyonic Giant Magnons in $\RP^3$}
Dyonic giant magnons in $S^3$ can be embedded into $\RP^3$.
The embedded solution is given by
\beq
\bm Z_{\rm em} =
\begin{pmatrix}
e^{it/2}(\cos\frac p2+i\sin\frac p2\tanh \frac u2)\\
e^{iv/2}\sin \frac p2{\rm sech} \frac u2\\
e^{-iv/2}\sin \frac p2{\rm sech} \frac u2\\
e^{-it/2}(\cos\frac p2-i\sin\frac p2\tanh \frac u2)
\end{pmatrix}\,,\label{eq:magrp3}
\eeq
where $u$ and $v$ are given by (\ref{eq:uv}).
We show that this solution is also obtained from our two-soliton solutions in Section \ref{sec:3}. 
In the solution \eqref{eq:2solABz1}-\eqref{eq:2solABz23}, we fix the parameters $\x_1=\x_2=\x$, $\bm \W^\dagger_1 \bm \W_2=0$, $\bm \W^\dagger_1 \bm \W_1=\bm \W^\dagger_2 \bm \W_2=1$.
Then we obtain
\beq\begin{split}
Z''_1 &= K_{11}(e^{u/2}+e^{-u/2})\frac{\x e^{u/2}+\bar\x e^{-u/2}}{\x-\bar \x}e^{it/2}, \\
Z''_4 &= K_{11}(e^{u/2}+e^{-u/2})\frac{\bar\x e^{u/2}+\x e^{-u/2}}{\x-\bar \x}e^{-it/2}, \\
\begin{pmatrix}
Z''_2 \\ Z''_3
\end{pmatrix}
&= K_{11}(e^{u/2}+e^{-u/2})\left[e^{iv/2}\bm\W_1+e^{-iv/2}\bm\W_2\right].\label{eq:rp3bydre}
\end{split}\eeq
Rescaling the coordinates by overall factor
\beq
\left(\frac{K_{11}(e^{u/2}+e^{-u/2})^2}{\x-\bar\x}\right)^{-1},
\eeq
we get
\beq\begin{split}
\tilde{Z}''_1 &= \left(\cos\frac p2+i\sin\frac p2\tanh \frac u2\right)e^{it/2}, \\
\tilde{Z}''_4 &= \left(\cos\frac p2-i\sin\frac p2\tanh \frac u2\right)e^{-it/2}, \\
\begin{pmatrix}
\tilde{Z}''_2 \\ \tilde{Z}''_3
\end{pmatrix}
&= \left(\sin\frac p2{\rm sech} \frac u2\right)(i\bm\W_1e^{iv/2}+i\bm\W_2e^{-iv/2}).
\end{split}\eeq
Thus we obtain the $\RP^3$ dyonic solution by choosing $\bm\W_{i}$ ($i=1,2$) as follows,
\beq
i\bm\W_1 = \begin{pmatrix}1\\0\end{pmatrix}, \ \ i\bm\W_2 = \begin{pmatrix}0\\1\end{pmatrix}\,.
\eeq
The above result means that the $\RP^3$ dyonic giant magnons are composites of two different kinds of $\CP^2$ dyonic giant magnons with equal soliton momenta.%
\footnote{Note that this was pointed out in \cite{Hollowood:2009sc}.}
On the gauge theory side, the composite operators of odd-site and even-site excitations with equal magnon momenta and $R$-charges correspond to these dyonic giant magnons.
This is consistent with the results of \cite{Bombardelli:2008qd, Lukowski:2008eq, Ahn:2008wd}.  

\subsection{Breather-like Solutions}
The breather-like solutions \eqref{eq:1solAB} can also be obtained from the two-soliton solutions.
In \eqref{eq:2solABz1}-\eqref{eq:2solABz23}, we fix the parameters $\x_1=\x_2=\x$, $\bm\W_1=\bm\W_2=\bm\W$, then get
\beq\begin{split}
Z''_1 &= K_{11}\left[\frac{\x(\bar\x^2-1)}{(\x-\bar\x)(\x\bar\x-1)}\frac1{\W^2}+\frac{\x e^u+\bar\x e^{-u}}{\x-\bar\x}+\frac{\x\bar\x e^{iv}+e^{-iv}}{\x\bar\x-1}\right]e^{it/2}, \\
Z''_4 &= K_{11}\left[\frac{\x(\bar\x^2-1)}{(\x-\bar\x)(\x\bar\x-1)}\frac1{\W^2}+\frac{\bar\x e^u+\x e^{-u}}{\x-\bar\x}+\frac{e^{iv}+\x\bar\x e^{-iv}}{\x\bar\x-1}\right]e^{-it/2}, \\
\begin{pmatrix}
Z''_2 \\ Z''_3
\end{pmatrix}
&= K_{11}\left[e^{(u-iv)/2}+e^{-(u-iv)/2}\right]\frac{\bm\W}{\W^2},
\end{split}\eeq
where $\W^2=\bm \W^\dagger \bm \W$.
We rescale the coordinates by overall factor $K_{11}^{-1}$ and replace the parameters as follows,
\begin{align}
\xi \to \bar\xi\,,\;\; \bm \W \to -\frac{\bm\W}{\W^2}\,,\;\; (u,v) \to (-u,v)\,.
\end{align}
The solution become
\beq\begin{split}
\tilde{Z}''_1 &= \left[-K_{11}-\frac{\x e^u+\bar\x e^{-u}}{\x-\bar\x}+\frac{\x\bar\x e^{iv}+e^{-iv}}{\x\bar\x-1}\right]e^{it/2}, \\
\tilde{Z}''_4 &= \left[-K_{11}-\frac{\bar\x e^u+\x e^{-u}}{\x-\bar\x}+\frac{e^{iv}+\x\bar\x e^{-iv}}{\x\bar\x-1}\right]e^{-it/2}, \\
\begin{pmatrix}
\tilde{Z}''_2 \\ \tilde{Z}''_3
\end{pmatrix}
&= -\left[e^{(u+iv)/2}+e^{-(u+iv)/2}\right]\bm\W.
\end{split}\eeq 
This is precisely identical to \eqref{eq:1solAB}.
The parameter configuration makes the solution that corresponds to the even-odd excitations like $Y^1Y_4^\dagger \rightarrow Y^2Y_2^\dagger$.
Thus it seems to be natural that the breather-like solution is reproduced by it.

\subsection{Reduction to Elementary Giant Magnons}
Let us consider the limit such that dyonic giant magnons reduce to elementary giant magnons.
If we take the limit $|\xi| \to 1$ in $\RP^3$ dyonic solution \eqref{eq:magrp3}, we obtain
\beq
\bm Z_{\rm em} = \begin{pmatrix}
e^{it}(\cos\frac p2+i\sin\frac p2\tanh \frac u2)\\
\sin \frac p2{\rm sech} \frac u2\\
\sin \frac p2{\rm sech} \frac u2\\
e^{-it}(\cos\frac p2-i\sin\frac p2\tanh \frac u2)
\end{pmatrix}.\label{eq:magrp2}
\eeq
This solution is the giant magnon in $\RP^2$.

To obtain giant magnons in $\CP^1$, we should be more careful.
Taking the limit $|\xi| \to 1$ naively in $\CP^2$ dyonic solution \eqref{eq:1solB}, the solution diverges.
To obtain the regular solution, we impose the condition such that
\begin{align}
|\x|\rightarrow1, \ \ \bm\W\rightarrow0, \ \ \frac{\bm\W^\dagger \bm\W}{\x\bar\x-1}\rightarrow \w,
\end{align}
where $\w$ is an arbitrary real parameter.
In this limit, \eqref{eq:1solB} reduces to
\beq
\bm Z' = \begin{pmatrix}
\left(-\frac{\x e^u}{\x-\bar\x}-\w\right)e^{it/2} \\
0 \\
0 \\
\left(-\frac{\bar\x e^u}{\x-\bar\x}-\w\right)e^{-it/2}
\end{pmatrix}. \label{eq:1solBlim}
\eeq
Multiplying $-(\x-\bar\x)e^{-u/2}$ by (\ref{eq:1solBlim}), we find
\beq
\tilde {\bm Z}' = \begin{pmatrix}
\left(\x e^{u/2}+\w(\x-\bar\x)e^{-u/2}\right)e^{it/2} \\
0 \\
0 \\
\left(\bar\x e^{u/2}+\w(\x-\bar\x)e^{-u/2}\right)e^{-it/2}
\end{pmatrix}.
\label{eq:1solBlim2}
\eeq
If we set $\w(\x-\bar\x)=-i$ and rotate \eqref{eq:1solBlim2} by unitary matrix $U=\diag(e^{-ip/4},1,1,e^{ip/4})$, we finally obtain
\beq
U\tilde{\bm Z}' = \begin{pmatrix}
\left(e^{u/2+ip/4}-ie^{-u/2-ip/4}\right)e^{it/2} \\
0 \\
0 \\
\left(e^{u/2-ip/4}-ie^{-u/2+ip/4}\right)e^{-it/2}
\end{pmatrix}\,.
\label{eq:CP1GM}
\eeq
It is easy to check that \eqref{eq:CP1GM} is the giant magnon in $\CP^1$.

\section{Boundstate $S$-matrix via Fusion}\label{app:S-bound}
In this appendix, we derive the boundstate $S$-matrices \eqref{eq:S^I}-\eqref{eq:S^IV}.
Let us start with the $S$-matrices for fundamental scattering.
The $SU(2|2)$ matrix $\hat{\mathbb{S}}(x_1^\pm,x_2^\pm)$ in \eqref{eq:S^OO},\eqref{eq:S^OE} can be written in terms of its matrix elements following the notation of \cite{Arutyunov:2006yd} (see also \cite{Beisert:2005tm}):
\begin{align}
\hat{\mathbb{S}}(x_1^\pm,x_2^\pm) =\sum_{i,j,k,l=1}^{4} S_{jl}^{ik}(x_1^\pm,x_2^\pm) E_i^j \otimes E_k^l\,,
\end{align}
where $E_i^j$ are the usual matrix unities.
For the cases (I)-(IV), the fundamental $S$-matrices are given by
\begin{align}
s^{(\rm I)}(x_1^\pm,x_2^\pm)&=S_0(x_1^\pm, x_2^\pm) S_{11}^{11}(x_1^\pm, x_2^\pm)
=s_{\rm BDS}(x_1^\pm,x_2^\pm) \sigma(x_1^\pm,x_2^\pm)\,,\\
s^{(\rm II)}(x_1^\pm,x_2^\pm)&=S_0(x_1^\pm, x_2^\pm) S_{12}^{12}(x_1^\pm, x_2^\pm)
=s_{\rm BDS}(x_1^\pm,x_2^\pm)s_2(x_1^\pm,x_2^\pm) \sigma(x_1^\pm,x_2^\pm)\,,\\
s^{(\rm III)}(x_1^\pm,x_2^\pm)&=\tilde{S}_0(x_1^\pm, x_2^\pm) S_{11}^{11}(x_1^\pm, x_2^\pm)
=\sigma(x_1^\pm,x_2^\pm)\,,\\
s^{(\rm IV)}(x_1^\pm,x_2^\pm)&=\tilde{S}_0(x_1^\pm, x_2^\pm) S_{12}^{12}(x_1^\pm, x_2^\pm)
=s_2(x_1^\pm,x_2^\pm) \sigma(x_1^\pm,x_2^\pm)\,,
\end{align}
where $S_{11}^{11}(x_1^\pm, x_2^\pm)$ and $S_{12}^{12}(x_1^\pm, x_2^\pm)$ take the forms
\begin{align}
S_{11}^{11}(x_1^\pm, x_2^\pm)&=\frac{x_1^+ - x_2^-}{x_1^- - x_2^+} \frac{\eta_1 \eta_2}{\tilde{\eta}_1 \tilde{\eta_2}}\,,\\
S_{12}^{12}(x_1^\pm, x_2^\pm)&=\left( \frac{x_1^+ - x_2^-}{x_1^- - x_2^+}+\frac{(x_1^--x_1^+)(x_2^--x_2^+)(x_1^++x_2^-)}{(x_1^--x_2^+)(x_1^-x_2^--x_1^+x_2^+)}\right)  \frac{\eta_1 \eta_2}{\tilde{\eta}_1 \tilde{\eta_2}}\,, \\
&=\frac{x_1^+ - x_2^+}{x_1^- - x_2^+} \cdot \frac{1-1/(x_1^- x_2^+)}{1-1/(x_1^- x_2^-)}\frac{\eta_1 \eta_2}{\tilde{\eta}_1 \tilde{\eta_2}}\,,
\end{align}
and $s_{\rm BDS}(x_1^\pm,x_2^\pm)$ and $s_2(x_1^\pm,x_2^\pm)$ are given by
\begin{align}
s_{\rm BDS}(x_1^\pm,x_2^\pm) &= \frac{x_1^+-x_2^-}{x_1^--x_2^+}\cdot \frac{1-1/(x_1^+ x_2^-)}{1-1/(x_1^+ x_2^-)}\,,\\
s_2 (x_1^\pm, x_2^\pm) &= \frac{x_1^+-x_2^+}{x_1^+-x_2^-}\cdot \frac{1-1/(x_1^- x_2^+)}{1-1/(x_1^- x_2^-)}\,.
\end{align}
The functions $\eta_i$, $\tilde\eta_i$ $(i=1,2)$ are gauge dependent parameters.
In spin-chain description, we should choose these parameters as
\begin{align}
\frac{\eta_1}{\tilde{\eta}_1}=\frac{\eta_2}{\tilde{\eta}_2}=1\,.
\label{eq:sc-frame}
\end{align}

Now let us consider the scattering of two magnon boundstates with charges (or magnon numbers) $Q_1$ and $Q_2$.
The spectral parameters of boundstates should satisfy the BPS pole conditions:
\begin{align}
x_{1,j_1}^-&=x_{1,j_1+1}^+\,, \quad (j_1=1,\dots,Q_1-1)\,, \label{eq:boundstate1}\\
x_{2,j_2}^-&=x_{2,j_2+1}^+\,, \quad (j_2=1,\dots,Q_2-1)\,. \label{eq:boundstate2}
\end{align}
For the convenience, we denote the outermost variables as $X_1^\pm,\,X_2^\pm$:
\begin{align}
X_1^+ \equiv x_{1,1}^+ \,,\;\; X_1^- \equiv x_{1,Q_1}^-\,,\;\;
X_2^+ \equiv x_{2,1}^+ \,,\;\; X_2^- \equiv x_{2,Q_2}^-\,.
\end{align}
The $S$-matrix of two boundstates can be constructed by the fusion procedure \cite{Chen:2006gq, Roiban:2006gs}.
The boundstate $S$-matrix is given by the product of $S$-matrices of two fundamental excitations:
\begin{align}
S^{(X)}(X_1^\pm,X_2^\pm) \equiv \prod_{j_1=1}^{Q_1}\prod_{j_2=1}^{Q_2} s^{(X)}(x_{1,j_1}^\pm, x_{2,j_2}^\pm)\,\;\;\;
(X={\rm I, II, III, IV})\,.
\end{align}
Using the boundstate conditions \eqref{eq:boundstate1} and \eqref{eq:boundstate2}, the product of the BDS parts becomes
\begin{align}
&S_{\rm BDS}(X_1^\pm, X_2^\pm) \equiv \prod_{j_1=1}^{Q_1}\prod_{j_2=1}^{Q_2} s_{\rm BDS}(x_{1,j_1}^\pm, x_{2,j_2}^\pm)\nonumber \\
& = \frac{(X_1^+ - X_2^-)(1-\frac{1}{X_1^+ X_2^-})(X_1^- - X_2^-)(1-\frac{1}{X_1^- X_2^-})}{(X_1^- - X_2^+)(1-\frac{1}{X_1^- X_2^+})(X_1^+ - X_2^+)(1-\frac{1}{X_1^+ X_2^+})}
\left(
\prod_{k=1}^{Q_1-1} \frac{X_1^++\frac{1}{X_1^+}-X_2^--\frac{1}{X_2^-}-\frac{ik}{h(\l)}}{X_1^++\frac{1}{X_1^+}-X_2^+-\frac{1}{X_2^+}-\frac{ik}{h(\l)}}
\right)^2\,.
\label{eq:S_BDS}
\end{align}
Similarly,
\begin{align}
S_{2}(X_1^\pm, X_2^\pm) &\equiv \prod_{j_1=1}^{Q_1}\prod_{j_2=1}^{Q_2} s_{2}(x_{1,j_1}^\pm, x_{2,j_2}^\pm)\nonumber  \\
& = \frac{(X_1^+ - X_2^+)(1-\frac{1}{X_1^- X_2^+})}{(X_1^+ - X_2^-)(1-\frac{1}{X_1^- X_2^-})}
\prod_{k=1}^{Q_1-1} \frac{X_1^++\frac{1}{X_1^+}-X_2^+-\frac{1}{X_2^+}-\frac{ik}{h(\l)}}{X_1^++\frac{1}{X_1^+}-X_2^--\frac{1}{X_2^-}-\frac{ik}{h(\l)}}\,.
\label{eq:S_2}
\end{align}
The product of the dressing parts has the same form as the fundamental case.
Combining all of these results, we obtain the boundstate $S$-matrices for the four types of scattering (I)-(IV), given by \eqref{eq:S^I}-\eqref{eq:S^IV}.


\bibliographystyle{utcaps}
\bibliography{refs}

\end{document}